\documentclass[superscriptaddress,showpacs,prc,twocolumn]{revtex4}
\usepackage{graphicx}
\usepackage{dcolumn}

\newcounter{subfigure}
\makeatletter
\renewcommand{\fnum@figure}{\figurename~\thefigure\alph{subfigure}}
\makeatother

\newcommand{\mc}[1]{\multicolumn{1}{c}{#1}}
\newcolumntype{x}[1]{D{.}{.}{#1}}
\begin{document}
\title{OVERTONES OF ISOSCALAR GIANT RESONANCES IN MEDIUM-HEAVY AND HEAVY NUCLEI}
\author{M.L. Gorelik}
\author{I.V. Safonov}
\affiliation{Moscow Engineering Physics Institute (State University), 115409
             Moscow,  Russia}
\author{M.H. Urin}
\email{urin@theor.mephi.ru} 
\affiliation{Moscow Engineering Physics Institute (State University), 115409
             Moscow,  Russia}
\affiliation{Kernfysisch Versneller Institute, 9747 AA Groningen,
             The Netherlands}
\date{\today}
\begin{abstract}
A semi-microscopic approach based on both the
continum-random-phase-approximation (CRPA) method and a phenomenological
treatment of the spreading effect is extended and applied to describe the main
properties (particle-hole strength distribution, energy-dependent transition
density, partial direct-nucleon-decay branching ratios) of the isoscalar giant
dipole, second monopole, and second quadrupole resonances. Abilities of the
approach are checked by description of gross properties of the main-tone
resonances. Calculation results obtained for the resonances in a few singly-
and doubly-closed-shell nuclei are compared with available experimental data.
\end{abstract}
\maketitle

\section{\label{sec:intro}INTRODUCTION}
Experimental and theoretical studies of high-energy giant resonances (GRs)
have been undertaken in recent years to understand better how the different
characteristics associated with GR formation (concentration of the
particle-hole strength, coupling to the continuum, the spreading effect) are
affected with increasing the GR energy. Many of known high-energy GRs are the
next vibration modes (the overtones) relative to the respective low-energy
GRs (the main tones). The lowest energy overtone is the isoscalar giant dipole
resonance (ISGDR) experimentally studied in a few medium-heavy and heavy
nuclei via the $(\alpha,\alpha')$ reaction \cite{Hara,Clark,Itoh,Nayak}.
The ISGDR is the overtone of the $1^{-}$ zero-energy spurious state (SS),
associated with center-of-mass motion. The experimental results of
Refs.~\cite{Hara,Clark,Itoh,Nayak} are concerned with distribution of the
respective dipole strength. Only recently direct nucleon decays of the ISGDR
have been firstly observed using the $(\alpha,\alpha'N)$ reactions
\cite{Hun1,Hun2,Garg}. These studies are planned to be continued \cite{Fuji}.
The isovector giant charge-exchange (in the $\beta^{-}$ channel) monopole and
spin-monopole resonances are the overtones of the isobaric analogue and
Gamow-Teller resonances, respectively. These overtones have been studied via
charge-exchange reactions \cite{Erell,Zeg}. Direct proton decays of the giant
spin-monopole resonance have been recently observed using the ($^3$He,$tp$)
reaction \cite{Zeg}. Other candidates for studies of high-energy GRs are the
overtones of the isoscalar giant monopole and quadrupole resonances (ISGMR2
and ISGQR2, respectively). The corresponding main tones, having relatively low
energy, have been much experimentally studied \cite{Hara,Clark,Itoh}.
However, the evidence for existence of the ISGQR2 has apparently been found
only recently \cite{Hun1,Hun2}.

Theoretical studies of the isoscalar overtones deal primarily with the ISGDR.
Microscopically, this GR is mainly due to $3\hbar\omega$
particle-hole-type excitations, while the overtones of the isoscalar quadrupole
and monopole GRs are mainly due to $4\hbar\omega$ excitations.
First microscopic calculations of the strength distribution for the ISGDR and
for all the above-mentioned isoscalar overtones have apparently been done in
the eighties (Refs.~\cite{Giai} and \cite{Mura1}, respectively). In recent
years Hartree-Fock + RPA calculations with the use of the Skyrme interactions
have been done to specify the ISGDR energy and, after comparing with
experimental data, to make a conclusion about the nuclear incompressibility
\cite{Colo,Shlom1}. Similar goals are pursued in the approaches based on the
relativistic version of the RPA and on the semi-classical treatment of nuclear
vibrations (see, e.g., Refs.~\cite{Vret} and \cite{Kolom}, respectively).
However, the microscopic structure and ``differential'' properties of
high-energy GRs, corresponding to collective excitations of nuclei as
finite-size open Fermi-systems, are of particular interest. Attempts to
describe the main properties (the strength distribution, energy-dependent
transition density, partial direct-nucleon-decay branching ratios) of the ISGDR
have been undertaken in Refs.~\cite{Gor1,Gor2} within a semi-microscopic
approach, based on both the CRPA method and a phenomelogical treatment of the
spreading effect. In these references the gross properties (parameters of the
strength distribution, transition density) have been satisfactorily described
for the ISGDR in a few medium-heavy and heavy nuclei. Contrary to
the gross properties, the partial branching ratios for direct nucleon decay
of a particular GR carry information on its microscopic structure and also on
coupling to the continuum and the spreading effect. The attempts (undertaken
in Refs.~\cite{Gor1,Gor2} for the ISGDR in $^{208}$Pb) to evaluate the partial
direct-nucleon decay branching ratios met with difficulties concerned with
taking the spreading effect into account. The way to overcome these
difficulties has been drawn in Ref.~\cite{Gor3}. The overtone of the isoscalar
monopole resonance was theoretically studied within a CRPA-based approach in
Ref.~\cite{Mura2} mainly for searching for narrow (``trapped'') resonances,
having a small relative strength. The gross properties of the ISGMR2 are
described in Ref.~\cite{Shlom2}, where the microscopic Hartree-Fock + RPA
approach with the use of the Skyrme interactions and also a semi-classical
approach are employed. Within the semi-microscopic approach, the main
properties of this resonance are briefly described in Ref.~\cite{Gor3}.

Motivated by both aspirations to describe the main properties of
high-energy GRs and forthcoming experimental data concerned with the isoscalar
overtones, we pursue the following goals in the present work:

\begin{enumerate}
\item extension of the CRPA-based semi-microscopic approach to describe direct
nucleon decay of high-energy GRs;

\item description of the gross properties of the isoscalar main-tone resonances
to check abilities of the approach and to find out probing operators
appropriate for overtone studies;

\item calculation of the partial direct-nucleon-decay branching ratios for the
high-energy component of the ISGDR;

\item description of the main properties of the second isoscalar giant monopole
and quadrupole resonances; and

\item comparison of the calculation results, obtained for singly- and
doubly-closed-shell nuclei $^{58}$Ni, $^{90}$Zr, $^{116}$Sn, $^{144}$Sm, and
$^{208}$Pb, with available experimental data.
\end{enumerate}

The paper is organized as follows. In Sect.~\ref{sec:ingred} the basic
elements, ingredients and new points of the approach are presented.
Sect.~\ref{sec:multipole} contains calculation results and available
experimental data on properties of the isoscalar giant main-tone and overtone
resonances. Discussion of the results and concluding remarks are given in
Sect.~\ref{sec:summary}.

\section{\label{sec:ingred}BASIC ELEMENTS AND INGREDIENTS OF THE APPROACH}
The continuum-RPA method and a phenomenological treatment of the spreading
effect are the basic elements of the semi-microscopic approach.
In implementations of the approach the following phenomenological input
quantities are used: a realistic nuclear mean field and the Landau-Migdal
partical-hole interaction bound together by selfconcistency conditions;
an energy- and radial-dependent smearing parameter.

\subsection{\label{sec:crpaeq}CRPA equations}
The CRPA equations are taken in the form accepted within the Migdal's finite
Fermi-system theory \cite{Migdal}. As applied to description of isoscalar
particle-hole-type excitations in closed-shell nuclei, these equations are
given in detail in Refs.~\cite{Gor1,Gor2} and not shown here.
(For reader's convenience we use (in the main) the notations of
Refs.~\cite{Gor1,Gor2} and sometimes refer to equations, tables and figures
from these references). The nuclear polarizibility $P_L(\omega)$, strength
function $S_L(\omega)$, and the energy-dependent transition density
$\rho_L(\vec r,\omega)=\rho_L(r,\omega)Y_{LM}(\vec n)$ ($\omega$ is
the excitation energy), corresponding to an isoscalar probing operator
$V_L(\vec r)=V_L(r)Y_{LM}(\vec n)$, can be calculated to describe gross
properties of the respective isoscalar GR within the CRPA. The above-listed
quantities are expressed via the radial part of the effective probing operators
$\tilde V_L^\alpha(r,\omega)$ ($\alpha=n,p$ is the isotopic index), which are
different from $V_L(r)$ due to core-polarization effects caused by the
particle-hole interaction. The $\omega$-dependence of the above quantities
comes from that of the free particle-hole propagator. The latter can be
expressed in terms of occupation numbers $n_\mu^\alpha$, radial bound-state
single-particle wave functions $\chi_\mu^\alpha$, and radial Green functions
$g_{(\lambda)}^\alpha(r,r',\epsilon_\mu \pm \omega)$ to take exactly the
single-particle continuum into account (see Eqs.~(1)--(3) of Ref.~\cite{Gor1}
and Eq.~(1) of Ref.~\cite{Gor2}).

When compared with Ref.~\cite{Migdal}, the new element of the CRPA equations
is used within the approach --- the direct-nucleon-escape amplitude
$M_c^L(\omega)$ (see Eqs.~(4), (5) of Ref.~\cite{Gor1}]).
This amplitude is proportional to the product of $(n_\mu^\alpha)^{1/2}$ and the
matrix element of the respective radial effective operator taken with the use
of the radial bound-state wave function $\chi_\mu^\alpha$ and the radial
continuum-state wave function $\chi_{\epsilon,(\lambda)}^{(+)\alpha}$
($\mu$ is the set of quantum numbers for an occupied single-particle level;
$\epsilon=\epsilon_\mu+\omega$ and $(\lambda)$ are the energy and quantum
numbers of an escaped nucleon, respectively, $c=\mu,(\lambda),\alpha$ is
the set of nucleon-decay-channel quantum numbers compatible with the respective
selection rules). The $\omega$-dependence of the direct-nucleon-escape
amplitude comes not only from the effective probing operator but also from
the continuum-state wave function.

The partial direct-nucleon-decay branching ratio $b_c^L(\delta)$ can be
reasonably defined as the ratio of the squared nucleon-escape amplitude
integrated over a certain excitation-energy interval
$\delta=\omega_1-\omega_2$ to the respective strength function integrated
over the same interval. The total branching ratio $b_{tot}^L=\sum_c b_c^L$ is
equal to unity for arbitrary interval $\delta$, as it follows from the unitary
condition, which is valid within the CRPA (Eq.~(4) of Ref.~\cite{Gor1}).

For comparison with the experimental branching ratios we sometimes replace the
occupation numbers $n_\mu$ in the expression for
$b_\mu^L=\sum_{(\lambda)} b_c^L$ with the respective experimental spectroscopic
factors $S_\mu$. In such a way we take phenomenologically into account coupling
of single-hole states $\mu^{-1}$ populated after direct nucleon decay of GRs to
low-energy collective states.

In description of the gross properties of high-energy GRs, nucleon pairing
in open-shell subsystems can be neglected with a high accuracy. To take into
account the effect of nucleon pairing on the direct-nucleon-decay branching
ratios in the CRPA equations we replace the occupation numbers $n_\mu$ with
the respective Bogolyubov's factors $v_\mu^2$. The latter can be calculated in
an isospin-selfconsistent way with the use of experimental pairing energies
\cite{Gor4}. To take phenomenologically into account coupling of
single-quasi-particle states populated after direct nucleon decay
to low-energy collective states the calculated $b_\mu$ value is divided by
$v_\mu^2$  and multiplied by $S_\mu$ \cite{Gor4}.

\begin{table}
\caption{\label{tab:I}The peak energy, total width (both in MeV),
and parameters $\eta_L$ (in fm$^2$) calculated for the $L=0,1,2$ isoscalar
GRs in nuclei under consideration. The respective experimental values
(given with errors) are taken from Ref.~\cite{Itoh} $(L=0,1)$
and Ref.~\cite{Young1} $(L=2)$.}
\begin{ruledtabular}
\begin{tabular}{rx{2.2}x{2.2}x{2.2}x{2.2}x{2.2}}
&\mc{$^{58}$Ni}&\mc{$^{90}$Zr}&\mc{$^{116}$Sn}&\mc{$^{144}$Sm}&\mc{$^{208}$Pb}\\
\hline
ISGMR&&&&&\\
$\omega_{peak}$&
 17.7 & 16.2            & 15.7            & 14.8            & 13.6            \\
&     &\mc{$16.9\pm0.1$}&\mc{$15.4\pm0.1$}&\mc{$15.4\pm0.1$}&\mc{$13.5\pm0.2$}\\
$\Gamma$&
 4.4 & 4.2            & 4.0            & 3.9            & 3.9            \\
&    &\mc{$4.3\pm0.1$}&\mc{$5.7\pm0.3$}&\mc{$3.9\pm0.2$}&\mc{$4.0\pm0.4$}\\
\hline
ISGQR&&&&&\\
$\omega_{peak}$&
 14.7 & 12.7            & 11.9            & 11.2            & 10.0            \\ 
&     &\mc{$14.0\pm0.2$}&\mc{$13.2\pm0.2$}&\mc{$12.2\pm0.2$}&\mc{$11.0\pm0.2$}\\
$\Gamma$&
 2.8 & 2.65           & 2.6            & 2.4            & 2.2            \\
&    &\mc{$3.4\pm0.2$}&\mc{$3.3\pm0.2$}&\mc{$2.4\pm0.2$}&\mc{$2.7\pm0.3$}\\
\hline
ISGDR&&&&&\\
$\eta_{L=1}$&
 22.71 & 29.28 & 34.74 & 39.74 & 51.07 \\
$\omega_{peak}^{LE}$&
9.2\footnote{The LE-component has additional maximum at 12.1 MeV}
       & 12.1  & 9.3   & 12.0  &
7.0\footnote{The LE-component has additional maximum at 11.4 MeV} \\
&             &\mc{$16.8\pm0.4$}&\mc{$15.5\pm0.4$}&\mc{$13.0\pm0.3$}&\mc{$13.0\pm0.2$}\\
$\omega_{peak}^{HE}$&
 27.4 & 26.0            & 26.1            & 25.0            & 22.9            \\
&     &\mc{$26.9\pm0.7$}&\mc{$24.9\pm0.7$}&\mc{$25.0\pm0.3$}&\mc{$22.8\pm0.3$}\\
\hline
ISGMR2&&&&&\\
$\eta_{L=0}$&
 42.04 & 52.75 & 55.57 & 54.50 &75.25 \\
$\omega_{peak}^{HE}$&
 30.8 & 29.9 & 32.0 & 33.8 & 32.1 \\
\hline 
ISGQR2&&&&&\\
$\eta_{L=2}$&
 24.38 & 30.50 & 35.90 & 40.41 & 52.54 \\
$\omega_{peak}^{HE}$&
 32.9 & 34.6 & 34.1 & 32.8 & 30.5 \\ 
\end{tabular}
\end{ruledtabular}
\end{table}

An important aspect of theoretital studies of GR overtones within the RPA is
the choice of an appropriate probing operator. It is convenient to choose this
operator with the condition that the main tone is not being excited. In this
case the overtone exhausts most of the respective particle-hole
strength. As applied to description of the ISGDR, charge-exchange giant
monopole, and spin-monopole resonances the choice of appropriate probing
operators is discussed in Refs.~\cite{Gor1,Gor4}, and \cite{Mura3},
respectively. In particular, the radial part of the isoscalar second-order
dipole probing operator $V_{L=1}^{(2)}(r)=r^3-\eta_{L=1}r$ is used for
description of the ISGDR. Parameter $\eta_{L=1}$ is defined by the condition:
$\int\rho_{L=1}^{SS}(r)V^{(2)}_{L=1}(r)r^2 dr=0$, where
$\rho_{L=1}^{SS}(r)$ is the spurious-state transition density. To describe the
 properties of the second isoscalar giant monopole and quadrupole resonances
the radial part of the respective second-order probing operators is taken in
the form: $V_L^{(2)}(r)=r^4-\eta_Lr^2$. The parameters $\eta_L$ in this
expression are determined by the condition:
\begin{equation}
\int\rho_L(r,\omega_{peak})V^{(2)}_L(r)r^2 dr=0,
\label{elimpole}
\end{equation}
where $\rho_L(r,\omega_{peak})$ is the radial part of the energy-dependent
main-tone transition density taken at the peak energy of the main-tone
strength function.

\begin{table}
\caption{\label{tab:II} Paramaters of the ISGMR and ISGQR calculated for
a certain excitation-energy interval. All the parameters are given in MeV
except for $x$, which is given in \%.} 
\begin{ruledtabular}
\begin{tabular}{rcx{2.2}x{1.2}ccx{2.2}x{1.2}c}
&\multicolumn{4}{c}{ISGMR}&\multicolumn{4}{c}{ISGQR}\\
nucl. & $\omega_1-\omega_2$ & \mc{$\bar\omega$} & \mc{$\Delta$} & $x$ &
       $\omega_ 1-\omega_2$ & \mc{$\bar\omega$} & \mc{$\Delta$} & $x$ \\
\hline
$^{58}$Ni  & 12-31 & 18.21 & 3.06 & 89 & 10-20 & 14.89 & 1.90 & 70 \\
$^{90}$Zr  & 10-25 & 16.47 & 2.64 & 87 & 9-18  & 13.07 & 1.78 & 70 \\
$^{116}$Sn & 10-20 & 15.38 & 2.13 & 79 & 8-19  & 12.51 & 2.13 & 72 \\
$^{144}$Sm & 10-20 & 14.85 & 2.05 & 80 & 7-19  & 11.91 & 2.31 & 74 \\
$^{208}$Pb & 10-20 & 13.89 & 2.05 & 79 & 6-17  & 10.51 & 2.18 & 71 \\
\end{tabular}
\end{ruledtabular}
\end{table}

\subsection{\label{sec:smear}Smearing procedure}
Within the approach, the speading effect, i.e. coupling of particle-hole-type
doorway states to many-quasi-particle configurations, is phenomenologically taken into
account in terms of a proper smearing parameter. Somewhat different smearing
procedures are used for description of low- and high-energy GRs. The
above-mentioned $\omega$-dependent quantities calculated within the CRPA for a
low-energy (``subbarrier'') GR can be expanded in terms of the
Breit-Wigner-type doorway-state resonances nonoverlapped on their total escape
widths $\Gamma^\uparrow$. To take the spreading effect in evaluation of the
energy-averaged $\omega$-dependent quantities into account, the doorway-state
resonances are smeared independently of one another. The smearing procedure
comprises the replacement of $\Gamma^\uparrow$ by $\Gamma^\uparrow + I$, or
(what is the same) the replacement of $\omega$ by $\omega +iI/2$.
The parameter $I$, the mean doorway-state spreading width, is fitted to
reproduce the experimental total width of the considered GR in calculations of
the energy-averaged strength function. To calculate the energy-averaged
direct-nucleon-escape amplitudes it is also necessary to average
the potential-barrier penetrability in a proper way because the directly
escaped nucleons have relatively low (``subbarrier'') energies.
Thus, the energy-averaged transition density and partial direct-nucleon-decay
branching ratios can be evaluated within the approach without the use of any
free parameters. This method has been employed in Ref.~\cite{Gor1} to decribe
quantitatively direct nucleon decay of the ISGMR in a few nuclei. As applied to
direct nucleon decay of the ISGDR in $^{208}$Pb, the method results only in
qualitative description because the respective doorway-state resonances are
significantly overlapped \cite{Gor1}. The above-outlined smearing procedure
is based on a statistical assumption: after energy-averaging the doorway states
``decay'' into many-quasi-particle configurations independently of one another
\cite{Mura3}. Such an assumption seems to be reasonable in view of the
complexity of many-quasi-particle configurations at high excitation energies.

\begin{table}
\caption{\label{tab:III} Parameters of the ISGDR calculated for different
excitation-energy intervals. All the parameters are given in MeV
except for $x$, which is given in \%.} 
\begin{ruledtabular}
\begin{tabular}{rcx{2.2}x{1.2}ccx{2.2}x{1.2}c}
&\multicolumn{4}{c}{LE-ISGDR}&\multicolumn{4}{c}{HE-ISGDR}\\
nucl. & $\omega_1-\omega_2$ & \mc{$\bar\omega$} & \mc{$\Delta$} & $x$ &
        $\omega_1-\omega_2$ & \mc{$\bar\omega$} & \mc{$\Delta$} & $x$ \\
\hline
$^{58}$Ni  & 5-16  & 11.56 & 2.08 & 12 & 16-36 & 26.62 & 4.55 & 75 \\
           &       &       &      &    & 16-60 & 27.94 & 6.36 & 86 \\
$^{90}$Zr  & 11-18 & 13.78 & 2.10 & 10 & 18-32 & 25.42 & 3.48 & 68 \\
           &  5-16 & 11.32 & 2.19 & 11 & 16-40 & 25.94 & 4.74 & 81 \\
           &       &       &      &    & 16-60 & 26.69 & 6.14 & 86 \\
$^{116}$Sn & 11-18 & 13.91 & 1.99 & 11 & 18-32 & 25.09 & 3.23 & 67 \\
           &  5-15 & 10.50 & 2.36 & 13 & 15-35 & 24.78 & 4.22 & 75 \\
           &       &       &      &    & 15-60 & 25.91 & 6.13 & 84 \\
$^{144}$Sm & 5-15  & 10.74 & 2.22 & 12 & 15-35 & 24.20 & 4.04 & 77 \\
           &       &       &      &    & 15-60 & 25.22 & 5.92 & 85 \\
$^{208}$Pb & 8-15  & 11.18 & 1.89 & 14 & 15-24 & 20.73 & 2.39 & 42 \\
           & 5-15  &  9.90 & 2.55 & 17 & 15-35 & 23.02 & 3.90 & 74 \\
           &       &       &      &    & 15-60 & 23.96 & 5.83 & 81 \\
\end{tabular}
\end{ruledtabular}
\end{table}

As follows from CRPA calculations of the high-energy GR strength function, the
doorway-state resonances are overlapped. In such a case, the smearing procedure
comprises the replacement of $\omega$ by $\omega+iI/2$ in the CRPA equations.
This replacement implies, in fact, the use of the imaginary part of the
single-particle potential, $\mp (i/2)I(r,\omega)$, when the radial Green
functions and continuum-state wave functions are calculated. As a result,
the energy-averaged $\omega$-dependent quantities (strength function
$\bar S_L(\omega)$, transition density $\bar\rho_L(r,\omega)$,
direct-nucleon-escape amplitude $\bar M^L_c(\omega))$ can be calculated at once
with the use of a radial- and energy-dependent smearing parameter in the CRPA
equations. In accordance with the statistical assumption, the following
parameterization for $I$ is used: $I(r,\omega) = I(\omega)f_{ws}(r,R^*,a)$,
where $f_{ws}$ is the Woods-Saxon function taken with $R^* > R$ ($R$ and $a$
are the radius and diffuseness of the isoscalar part of the nuclear mean field,
respectively). The calculated strength distribution and transition density of
high-energy GRs are found to be almost independent of the ``cut off'' radius
at $R^*>(1.7-1.9)R$. The energy-dependent part of the smearing parameter
$I(\omega)$ is taken in the form:
\begin{figure}
\includegraphics[scale=0.9]{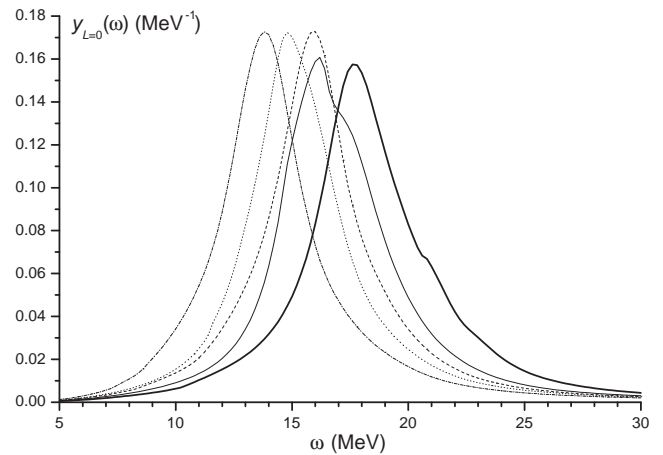}
\caption{The calculated relative energy-weighted strength function for
the ISGMR. The thick, thin, dashed, dotted, and dash-dotted lines are for
$^{58}$Ni, $^{90}$Zr, $^{116}$Sn, $^{144}$Sm, and $^{208}$Pb, respectively.}
\addtocounter{figure}{-1}
\addtocounter{subfigure}{1}
\end{figure}
\begin{figure}
\includegraphics[scale=0.9]{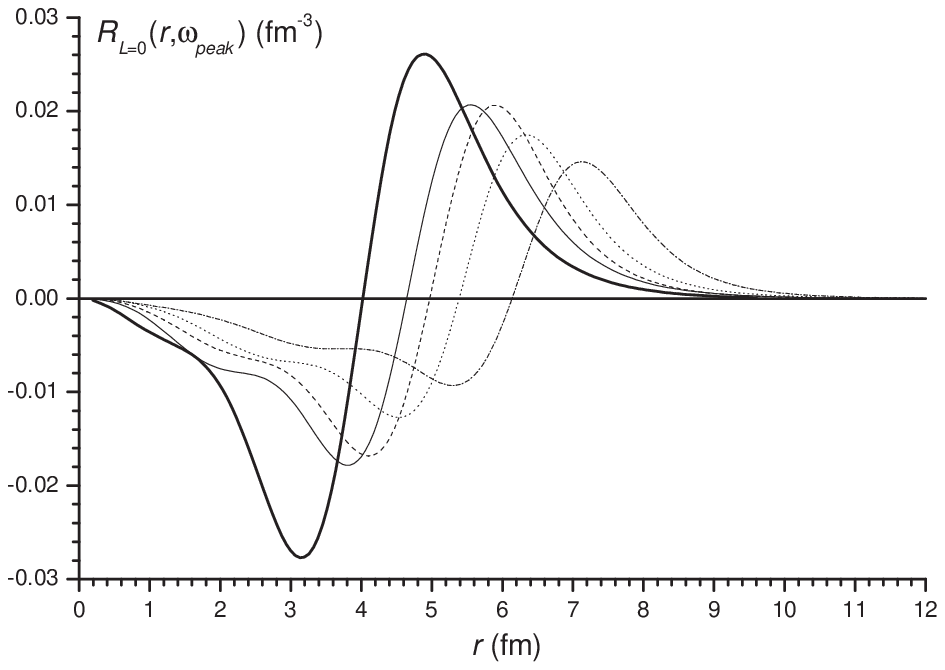}
\caption{The reduced energy-dependent transition density calculated at the peak
energy of the ISGMR (Table~\ref{tab:I}). The notations are the same as in FIG.~1a.}
\setcounter{subfigure}{1}
\end{figure}
\begin{table*}
\caption{\label{tab:IV}Comparison of calculated and experimental parameters
of the ISGMR and ISGDR. The experimental data (given with errors) are taken
from Refs.~\cite{Young2} and \cite{Clark}, respectively. All the parameters
are given in MeV.} 
\begin{ruledtabular}
\begin{tabular}{rccx{2.1}cx{2.1}cx{2.1}}
&&\multicolumn{2}{c}{$^{90}$Zr}&\multicolumn{2}{c}{$^{116}$Sn}
&\multicolumn{2}{c}{$^{208}$Pb}\\
\hline
ISGMR&
$\bar\omega$&$17.89\pm0.20$&16.5&$16.07\pm0.12$&15.4&$14.17\pm0.28$&13.9\\
&$\Delta$   &$3.14\pm0.09$ & 2.6&$2.16\pm0.08$ &2.1&$1.93\pm0.15$ &2.1\\
LE-ISGDR&
$\bar\omega$&$16.2\pm0.8$&13.8&$14.7\pm0.5$&13.9&$12.2\pm0.6$&11.2\\
&$\Delta$   &$1.9\pm0.7$ & 2.1&$1.6\pm0.5$ & 2.0&$1.9\pm0.5$ & 1.9\\
HE-ISGDR&
$\bar\omega$&$25.7\pm0.7$&25.4&$23.0\pm0.6$&25.1&$19.9\pm0.8$&20.7\\
&$\Delta$   &$3.5\pm0.6$ & 3.5&$3.7\pm0.5$ & 3.2&$2.5\pm0.6$ & 2.4\\
\end{tabular}
\end{ruledtabular}
\end{table*}
\begin{eqnarray}
I(\omega)&=&\alpha\frac{(\omega-\Delta)^2}{1+(\omega-\Delta)^2/B^2},
\quad\omega>\Delta;\nonumber\\*
I(\omega)&=&0,\quad\omega<\Delta
\label{iomega}
\end{eqnarray}
with universal parameters. Such an energy dependence is used for the
absorption potential in some versions of the optical model for nucleon-nucleus
scattering \cite{Maha}. The above-outlined smearing procedure has been used in
Ref.~\cite{Gor2} to describe quantitatively the gross properties of the ISGMR
and ISGDR in a few nuclei. However, this procedure can not be directly applied
to evaluation of the energy-averaged escape amplitudes $\bar M^L_c(\omega)$
(and, therefore, the respective branching ratios) because of nonphysical
absorption of escaped nucleons outside the nucleus. For this reason, in
the present work all the energy-averaged $\omega$-dependent quantities are
calculated using the ``cut off'' radius $R^*=R$ together with the properly
increased value of the intensity $\alpha$ in Eq.~(\ref{iomega}). Such a
choice allows us to describe correctly the respective single-particle
resonances in the energy dependence of the continuum-state wave function,
while the calculated gross properties of a high-energy GR are found almost
the same within both versions of the smearing procedure.
In Refs.~\cite{Gor2,Gor4,Rodin} the branching ratios calculated for a few
overtones in $^{208}$Pb are overestimated because of ignoring absorption
of escaped nucleons inside the nucleus. This shortcoming is partially
eliminated in Ref.~\cite{Gor3}, where all $\omega$-dependent quantities are
calculated using the smearing parameter with intermediate radius $R^*=1.3R$
to take the spreading effect into account in evaluation of the branching
ratios for the ISGDR in a few nuclei.

\begin{figure}
\includegraphics[scale=0.9]{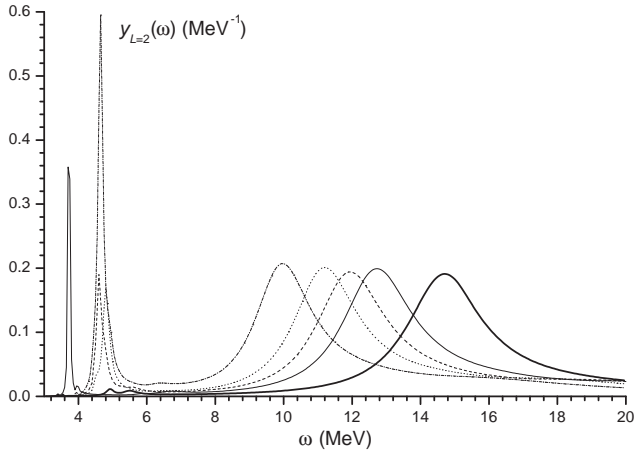}
\caption{The calculated relative energy-weighted strength function for the
ISGQR. The notations are the same as in FIG.~1a.}
\addtocounter{figure}{-1}
\addtocounter{subfigure}{1}
\end{figure}

In conclusion of this Subsect., we define the energy-averaged quantities
suitable for description of the GR main properties. The relative energy-weighted
strength function $y_L(\omega)=\omega\bar S_L(\omega)/(EWSR)_L$ is used to show
exhaustion of the respective energy-weighted sum rule (EWSR) by a particular GR.
The use of the reduced energy-dependent transition density
$R_L(r,\omega)=r^2\bar\rho_L(r,\omega)\bar S^{-1/2}_L(\omega)$ normalized by
the condition $\int R_L(r,\omega)V_L(r) dr=1$ is convenient to compare
the transition densities related to different energy regions~\cite{Gor2}.
The squared and properly normalized energy-averaged direct-nucleon-escape
amplitude determines the respective differential partial branching ratio:
\begin{equation}
\frac{d\bar b_\mu^L(\omega)}{d\omega}=
\frac{\displaystyle\sum_\lambda|\bar M_\lambda^L(\omega)|^2}
{\displaystyle\int_{\omega_1}^{\omega_2}\bar S_L (\omega) d\omega}.
\label{difbnu}
\end{equation}
These quantities show how the partial and total branching ratios are formed.

\begin{figure}
\includegraphics[scale=0.9]{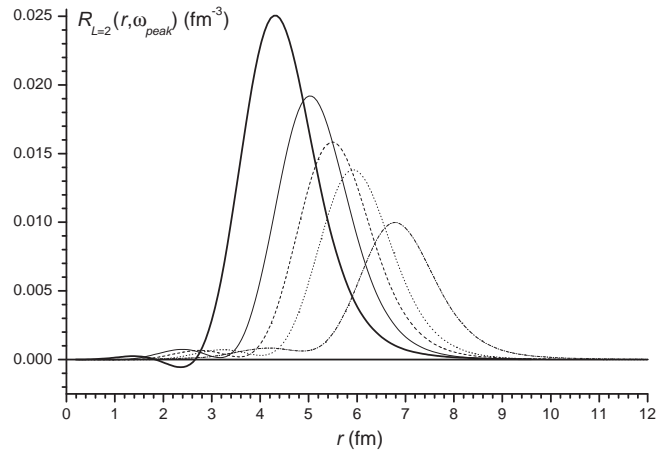}
\caption{The reduced energy-dependent transition density calculated at the peak
energy of the ISGQR (Table~\ref{tab:I}). The notations are the same as in FIG.~1a.}
\setcounter{subfigure}{1}
\end{figure}

\subsection{\label{sec:appingred}Ingredients of the approach}
Within the approach, a phenomenological isoscalar part of the nuclear mean
field (including the spin-orbit term) and the (momentum-independent)
Landau-Migdal particle-hole interaction are used as the input quantities for
CRPA calculations. Parameterization of these quantities is explicitly given
in Ref.~\cite{Gor1}. Within the RPA, the isospin symmetry of the model
Hamiltonian can be restored. As a result, the isovector part of the mean field
is calculated selfconsistently via the Landau-Migdal isovector parameter $f'$
and the neutron-excess density (see, e.g., Ref.~\cite{Gor4}). The mean
Coulomb field is calculated also selfconsistently via the proton density.

Within the CRPA, the distribution of the isoscalar dipole strength
(corresponding to the probing operator with the radial part $V_{L=1}(r)=r$)
can be calculated for a given model Hamiltonian. In studies of
Refs.~\cite{Gor1,Gor2}, the Landau-Migdal isoscalar parameter $f^{ex}$ is
chosen for each nucleus to make the energy of the $1^{-}$ spurious state
close to zero. If the translation invariance of the model Hamiltonian was
fully restored, the spurious state would exhaust 100\%
of the EWSR corresponding to the above-mentioned probing operator. Within the
current version of the approach, the SS exhausts in all cases more than 92\%
of this sum rule, the exact percentage being dependent on the
nucleus~\cite{Gor1,Gor2}. The small part of the spurious strength (less than 8\%)
is distributed mainly among isoscalar dipole $1\hbar\omega$ particle-hole-type
excitations.

\begin{figure}
\includegraphics[scale=0.9]{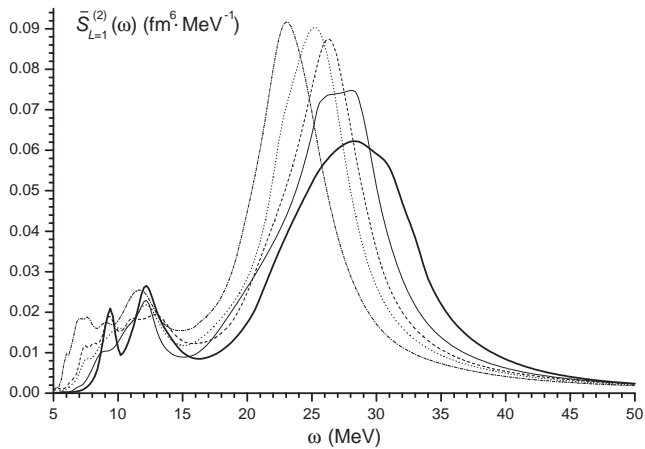}
\caption{The calculated energy-averaged strength function for the ISGDR.
The notations are the same as in FIG.~1a.}
\addtocounter{figure}{-1}
\addtocounter{subfigure}{1}
\end{figure}
\begin{table}
\caption{\label{tab:V} Comparison of calculated and experimental
parameters of the ISGMR and ISGQR in $^{58}$Ni. The experimental data are
taken from Ref.~\cite{Lui}. All the parameters are given in MeV.} 
\begin{ruledtabular}
\begin{tabular}{rccx{2.1}}
ISGMR & $\bar\omega$ & $20.30^{+1.69}_{-0.14}$ & 18.2 \\
      & $\Delta$     &  $4.25^{+0.69}_{-0.23}$ &  3.1 \\
\hline
ISGQR & $\bar\omega$ & $16.1\pm0.3$ & 14.9 \\
      & $\Delta$     &  $2.4\pm0.2$ &  1.9 \\
\end{tabular}
\end{ruledtabular}
\end{table}

The calculation results presented in Sect.~\ref{sec:multipole} are obtained
with the use of model parameters taken, in the main, from previous studies
of Refs.~\cite{Gor1,Gor2}. The mean-field parameters and the Landau-Migdal
isovector parameter $f'=1.0$ are taken from Ref.~\cite{Gor1}, where
the experimental nucleon separation energies have been satisfactorily
described for closed-shell subsystems in nuclei with $A=90-208$. The isoscalar
Landau-Migdal paramater $f^{in}$ is taken equal to $0.0875$ in agreement
with the systematics of Refs.~\cite{Migdal,Speth}, while the values
of parameter $f^{ex}=-(2.7-2.9)$ are found in Ref.~\cite{Gor2} in the
above-described way for each nucleus under consideration. The universal
parameters of Eq.~(\ref{iomega}) used for description of the spreading effect
$\alpha=0.125$ MeV$^{-1}$, $\Delta=3$ MeV, $B=7$ MeV are taken the
same, as in Ref.~\cite{Gor2}, except for the $\alpha$ value
($\alpha=0.085$ MeV$^{-1}$ is used in Ref.~\cite{Gor2} together with $R^*=1.8R$).

\section{\label{sec:multipole}Properties of the isoscalar giant multipole
resonances}
\subsection{\label{sec:main-tone}Gross properties of the main-tone resonances}
We start with description of the main-tone resonances. This description
allows us

\begin{itemize}
\item to check the quality of restoration of the translation invariance
of the model within the CRPA;

\item to check abilities of the semi-microscopic approach because the gross
properties of the isoscalar giant monopole and quadrupole resonances are
extensively studied experimentally;

\item to evaluate the parameters $\eta_L$ in the expression for the radial part
of the second-order probing operators appropriate for microscopic studies of
the overtones.
\end{itemize}

\begin{figure}
\includegraphics[scale=0.9]{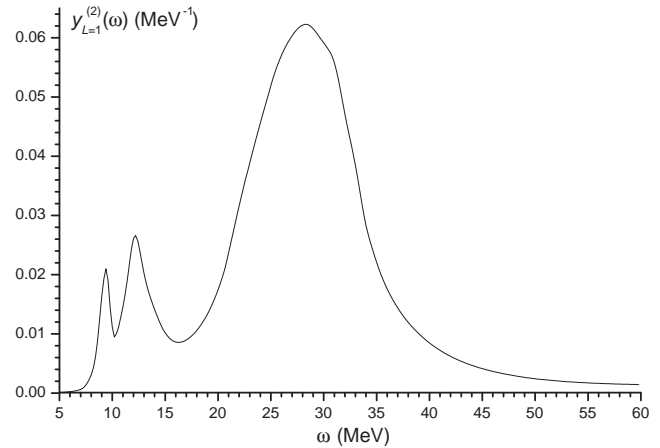}
\caption{The calculated relative energy-weighted strength function for the
ISGDR in $^{58}$Ni.}
\addtocounter{figure}{-1}
\addtocounter{subfigure}{1}
\end{figure}

The $1^-$ spurious state associated with the center-of-mass motion is
the lowest energy main-tone state. The method to find out properties of the SS
within the CRPA is described in detail in Ref.~\cite{Gor1} (see also
Subsect.~\ref{sec:appingred}). The method allows us to specify the value of
the Landau-Migdal parameter $f^{ex}$ and to calculate the characteristics of
the SS, relative isoscalar dipole strength $x^{SS}_{L=1}$ and the transition
density $\rho_{L=1}^{SS}(r)$. The calculated values of $f^{ex}$ and
$x^{SS}_{L=1}$ for nuclei under consideration (except for $^{58}$Ni) are given
in Table~I of Ref.~\cite{Gor2}. For $^{58}$Ni we obtain
$f^{ex}=-2.646$ and $x^{SS}_{L=1}=96$\%.
The values $f^{ex}=-2.789$ and $x^{SS}_{L=1}=94$\%
are obtained for $^{116}$Sn with taking neutron pairing into account. 
The radial dependence of the calculated transition density
$\rho_{L=1}^{SS}(r)$ is found to be close to $d\rho(r)/dr$, where the
ground-state density $\rho(r)$ is determined by the bound-state wave functions
for the occupied levels:
\begin{figure}
\includegraphics[scale=0.9]{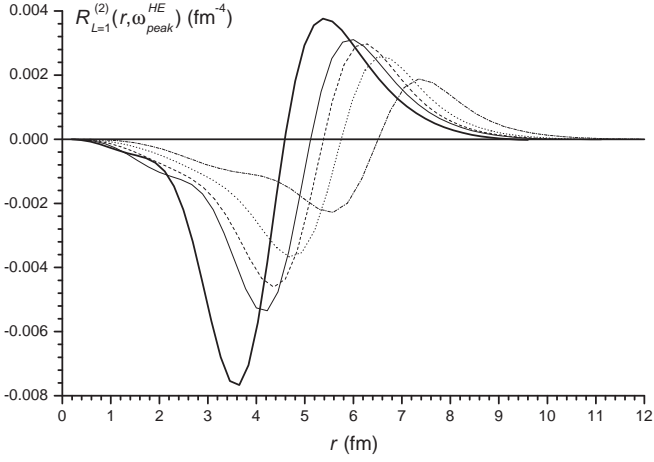}
\caption{The reduced energy-dependent transition density calculated at the peak
energy of the HE-ISGDR (Table~\ref{tab:I}). The notations are the same as in FIG.~1a.}
\addtocounter{figure}{-1}
\addtocounter{subfigure}{1}
\end{figure}
\begin{equation}
\rho(r)=\sum_{\alpha=n,p}\sum_\mu
\frac{(2j_\mu+1)}{4\pi r^2} n^\alpha_\mu
(\chi^\alpha_\mu (r))^2.
\label{grstrho}
\end{equation}
This result is quite expectable, because the calculated relative strength
$x^{SS}_{L=1}$ is rather close to 100\%.
For the same reason, the parameter $\eta_{L=1}$ in the expression for
the second-order dipole operator (Subsect.~\ref{sec:crpaeq}) is almost equal
to the value $5<r^2>/3$ (averaging is performed over the ground-state density
of Eq.~(\ref{grstrho})). This value is employed in many works
(see, e.g. Refs.~\cite{Giai,Colo,Shlom1,Gor1,Gor2}). The $\eta_{L=1}$ values used
in calculations of the main properties of the ISGDR in nuclei under
consideration are given in Table~\ref{tab:I}.

\begin{figure}
\includegraphics[scale=0.9]{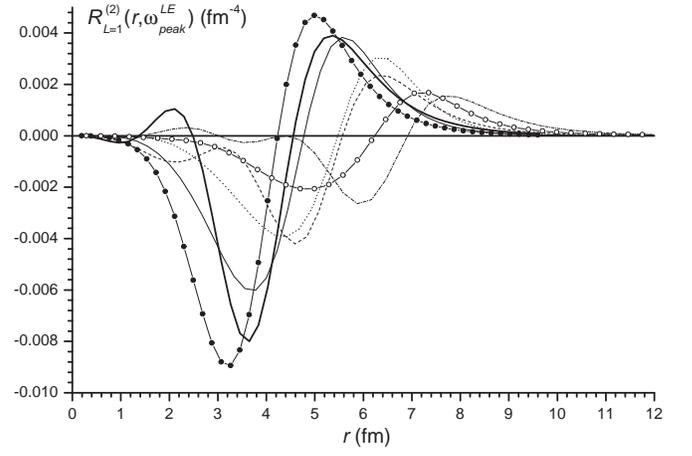}
\caption{The reduced energy-dependent transition density calculated at the peak
energy of the LE-ISGDR (Table~\ref{tab:I}). The notations are the same as
in FIG.~1a. Curves with filled circles and open circles correspond,
respectively, to the additional maximuma for $^{58}$Ni and $^{208}$Pb
as given in Table~\ref{tab:I}.}
\setcounter{subfigure}{0}
\end{figure}
\begin{figure}
\includegraphics[scale=0.9]{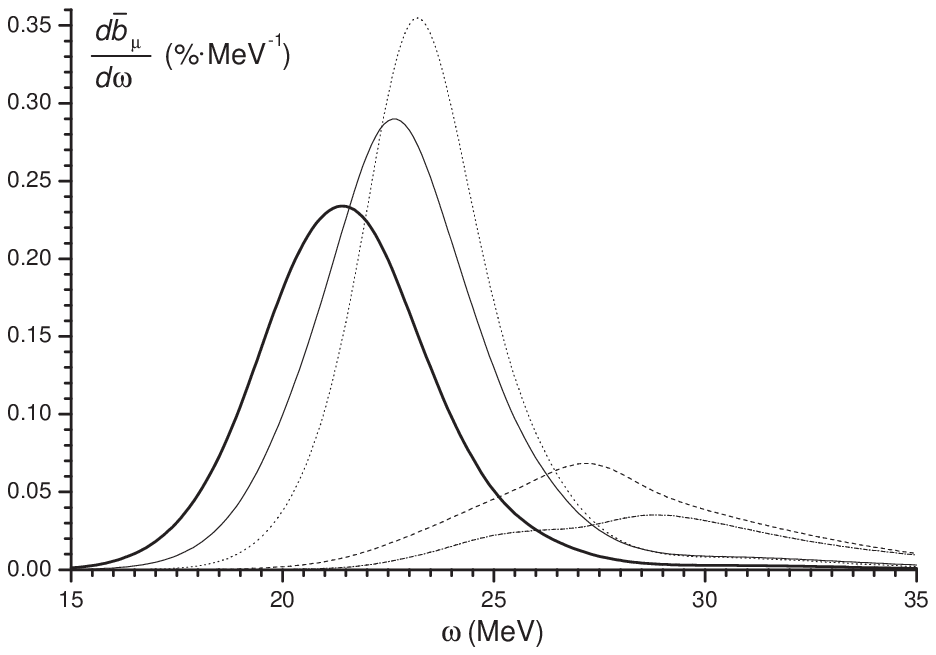}
\caption{Differential partial proton branching ratios for the HE-ISGDR
in $^{208}$Pb calculated for several decay channels $(S_\mu = 1)$.
The thick, thin, dashed, dotted and dash-dotted lines are for direct decay
into $3s_{1/2}$, $2d_{3/2}$, $1h_{11/2}$, $2d_{5/2}$, and $1g_{7/2}$ one-hole
states of $^{207}$Tl, respectively.}
\addtocounter{subfigure}{1}
\end{figure}

The isoscalar monopole and quadrupole GRs in nuclei under consideration are
experimentally studied in many works \cite{Hara,Clark,Itoh,Young1,Young2,Lui}.
The deduced strength distributions are presented either in terms of the peak
energy $\omega_{peak}^L$ and total width $\Gamma_L$ (obtained by the Lorentzian or
Gaussian fit), or in terms of the mean energy $\bar\omega_L$ and RMS energy
dispersion $\Delta_L$ (obtained for a certain excitation energy interval).
These data are used below for comparison with the results, obtained within
the semi-microscpopic approach. The energy-averaged strength functions
$\bar S_L(\omega)$ calculated with use of the radial part $V_L(r)=r^2\ (L=0,2)$
of the first-order probing operators exhibit well prominent resonances in their
energy dependence. The respective peak energies and total widths are
given in Table~\ref{tab:I} together with the corresponding experimental values.
The relative energy-weighted strength functions
$y_L(\omega)=\omega\bar S_L(\omega)/(EWSR)_L$ are shown in Fig.~1a $(L=0)$
and in Fig.~2a $(L=2)$ for nuclei under consideration. Microscopically
the ISGMR and ISGQR are mainly due to $2\hbar\omega$ particle-hole-type
excitations. The low-energy component of the ISGQR (Fig.~2a) is due to
$0\hbar\omega$ excitations corresponding to single-particle transitions with
changing both the radial and orbital quantum numbers and having, therefore,
a small relative strength. The calculated strength functions are used
to evaluate (for a certain energy interval) the parameters of the ISGMR
and ISGQR: mean energy $\bar\omega_L$, RMS energy dispersion $\Delta_L$,
relative strength $x_L$. These values are listed in Table~\ref{tab:II}.
Comparison with the available experimental data is given in Tables~\ref{tab:IV}
and \ref{tab:V}. The reduced transition densities $R_L(r,\omega_{peak}^L)$
calculated for the main-tone isoscalar monopole and quadrupole GRs in nuclei
under consideration are shown in Fig.~1b and Fig.~2b, respectively.
The transition densities are used to evaluate parameters $\eta_L$ accordingly
to Eq.~(\ref{elimpole}). These values are given in Table~\ref{tab:I}.

\begin{figure}
\includegraphics[scale=0.9]{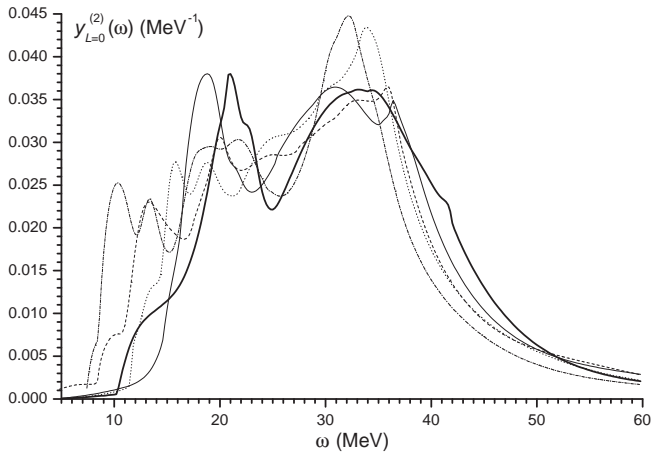}
\caption{The calculated relative energy-weighted strength function for the
ISGMR2. The notations are the same as in FIG.~1a.}
\addtocounter{figure}{-1}
\addtocounter{subfigure}{1}
\end{figure}

\subsection{\label{sec:isgdr}Properties of the ISGDR}
The gross properties of the ISGDR in nuclei under consideration are
described within the approach using the isoscalar second-order dipole
probing operator with parameters $\eta_{L=1}$ taken from Table~\ref{tab:I}.
The calculated energy-averaged strength function $\bar S_{L=1}^{(2)}(\omega)$
(Fig.~3a) exhibits a ``bi-modal'' energy dependence, corresponding to the
low- and high-energy components of the ISGDR (LE- and HE-ISGDR,
respectively). As a resonance in the strength function energy dependence,
the HE-ISGDR is well prominent and can be described in terms of the peak
energy and total width, while the LE-ISGDR is less prominent.
The peak energies of both components are given in Table~\ref{tab:I} together
with the latest experimental data. The relative energy-weighted strength
functions $y_{L=1}^{(2)}(\omega)$ are very close to those, shown in Fig.~1
of Ref.~\cite{Gor2} for nuclei under consideration (except for $^{58}$Ni)
and are not shown here. The parameters of both components calculated
with the use of the above strength functions are given in Table~\ref{tab:III}.
A part of these results related to certain excitation-energy intervals are
compared with the available experimental data deduced for the same intervals
(Table~\ref{tab:IV}). The strength function $y_{L=1}^{(2)}(\omega)$ calculated
for $^{58}$Ni is shown in Fig.~3b. Because the main-tone transition density is
nodeless, the energy-dependent transition density
$R_{L=1}^{(2)}(r,\omega_{peak})$ calculated at the peak energy of each
component exhibits one-node radial dependence (Figs.~3c,~3d).

\begin{figure}
\includegraphics[scale=0.9]{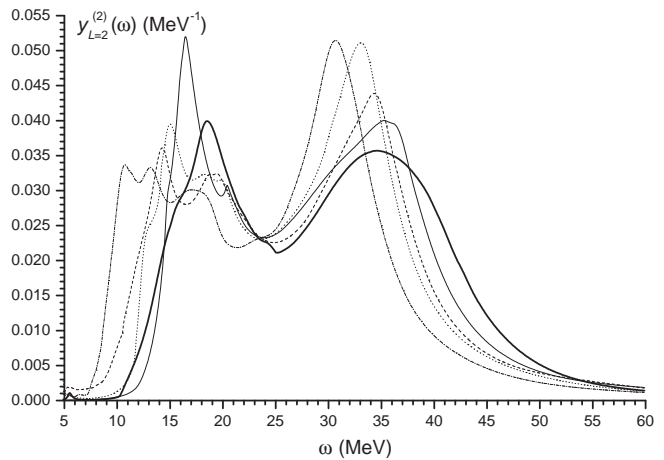}
\caption{The calculated relative energy-weighted strength function for the
ISGQR2. The notations are the same as in FIG.~1a.}
\setcounter{subfigure}{1}
\end{figure}
\begin{table}
\caption{\label{tab:VI} Calculated partial branching ratios for direct proton
decay of the HE-ISGRs in $^{58}$Ni $(S_\mu = 1)$. Branching ratios are given in \%.} 
\begin{ruledtabular}
\begin{tabular}{rx{2.1}x{2.1}x{2.1}}
$\mu^{-1}$ & \mc{$\bar b^{L=1}_\mu$} & \mc{$\bar b^{L=0}_\mu$} & \mc{$\bar b^{L=2}_\mu$} \\
           & \mc{(15-40 MeV)}        & \mc{(23-40 MeV)}        & \mc{(25-40 MeV)}   \\
\hline
(7/2)$^-$   & 16.4 & 12.5 & 19.7 \\
(1/2)$^+$   & 4.7  & 6.0  &  4.1 \\
(3/2)$^+$   & 5.4  & 7.0  &  5.8 \\
(5/2)$^+$   & 6.6  & 9.7  &  7.9 \\
(1/2)$^-$   & 0.8  & 1.5  &  1.0 \\
(3/2)$^-$   & 1.1  & 2.8  &  2.1 \\
\hline
$\bar b^{tot}_p$ & 35.0 & 39.7 & 40.6 \\
\end{tabular}
\end{ruledtabular}
\end{table}

Direct nucleon decay of giant resonances is closely related to their
microscopic structure. For this reason, the decay probabilities belong to 
the main properties of GRs together with the energy,
total width, and transition density. The use of the CRPA method together
with the phenomenological treatment of the spreading effect
allows us to evaluate the partial direct-nucleon-decay branching ratios for
various GRs within the semi-microscopical approach. Turning to the HE-ISGDR
in nuclei under consideration, we present the direct-nucleon-decay
branching ratios $\bar b_\mu^{L=1}$ calculated with the use
of unit spectroscopic factors $S_\mu$ for single-hole states in closed-shell
subsystems and of unit ratio $S_\mu/v^2_\mu$ for single-quasiparticle
states in open-shell subsystems (Tables~\ref{tab:VI}-\ref{tab:X}). 
Some partial branching ratios corresponding to population of deep-hole states
are not shown, but they are included in the values of the respective total
branching ratios also given in the tables. The recent experimental data of
Refs.~\cite{Hun2,Garg} on the partial direct-proton-decay branching ratios
for the HE-ISGDR in $^{208}$Pb are given in Table~\ref{tab:XI} together with
the respective calculated values obtained with the use of the experimental
spectroscopic factors. Calculated for the same resonance, the differential
partial proton branching ratios of Eq.~(\ref{difbnu}) are shown in Fig.~4.
One can see from this figure the role of the penetrability factor in formation
of the HE-ISGDR in different proton decay channels.

\begin{figure}
\includegraphics[scale=0.9]{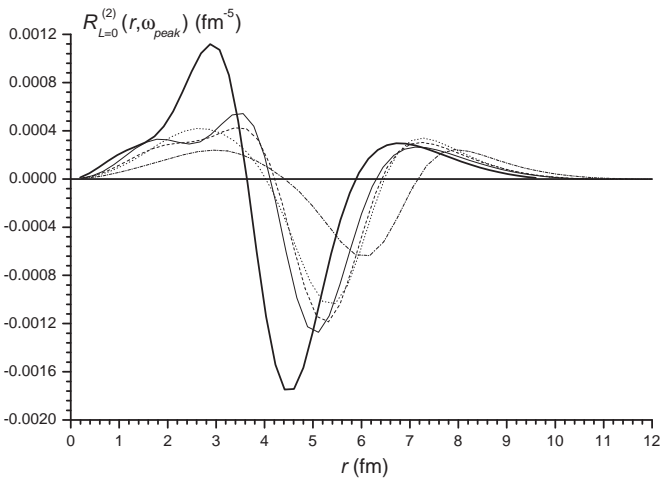}
\caption{The reduced energy-dependent transition density calculated at the peak
energy of the ISGMR2 (Table~\ref{tab:I}). The notations are the same as in FIG.~1a.}
\addtocounter{figure}{-1}
\addtocounter{subfigure}{1}
\end{figure}
\begin{figure}
\includegraphics[scale=0.9]{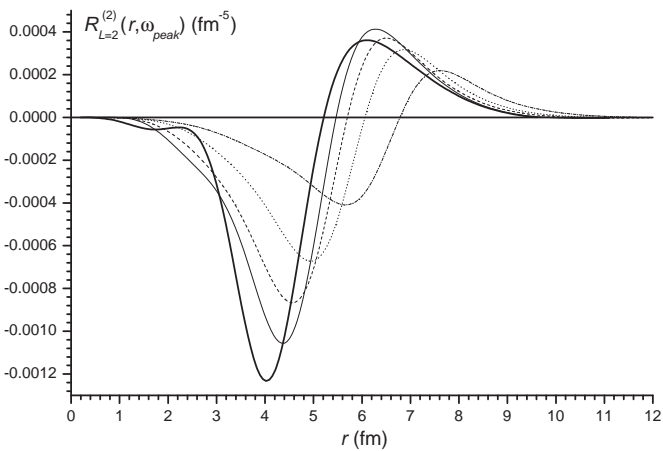}
\caption{The reduced energy-dependent transition density calculated at the peak
energy of the ISGQR2 (Table~\ref{tab:I}). The notations are the same as in FIG.~1a.}
\setcounter{subfigure}{0}
\end{figure}

\subsection{\label{sec:overtones}Main properties of the ISGMR2 and ISGQR2}
We describe the gross properties of the isoscalar monopole and
quadrupole overtones in nuclei under consideration, using the respective
second-order probing operators with parameters $\eta_L\ (L=0,2)$ taken from
Table~\ref{tab:I}. The overtone strength is distributed over a wide energy
interval exhibiting the main peak at a high excitation energy.
The $\omega_{peak}^L$ values are given in Table~\ref{tab:I}. The relative
energy-weighted strength functions $y_L^{(2)}(\omega)$ are shown in Figs.~5a
$(L=0)$ and 5b $(L=2)$, while the parameters of the high-energy components are
given in Table~\ref{tab:XII}. Because the main-tone transition density has one
node inside of the nucleus ($L=0$, Fig.~1a) or is nodeless ($L=2$, Fig.~1b),
the overtone transition density has two nodes ($L=0$, Fig.~6a) or one node
($L=2$, Fig.~6b), respectively. The direct-nucleon-decay branching ratios
$\bar b_{\mu}^L$, calculated for the high-energy components of both overtones
with the use of unit spectroscopic factors, are given in
Tables~\ref{tab:VI}-\ref{tab:X}. Partial branching ratios for direct proton
decay of the above resonances in $^{208}$Pb are also calculated with the use
of experimental spectroscopic factors (Table~\ref{tab:XI}).

\begin{figure}
\includegraphics[scale=0.9]{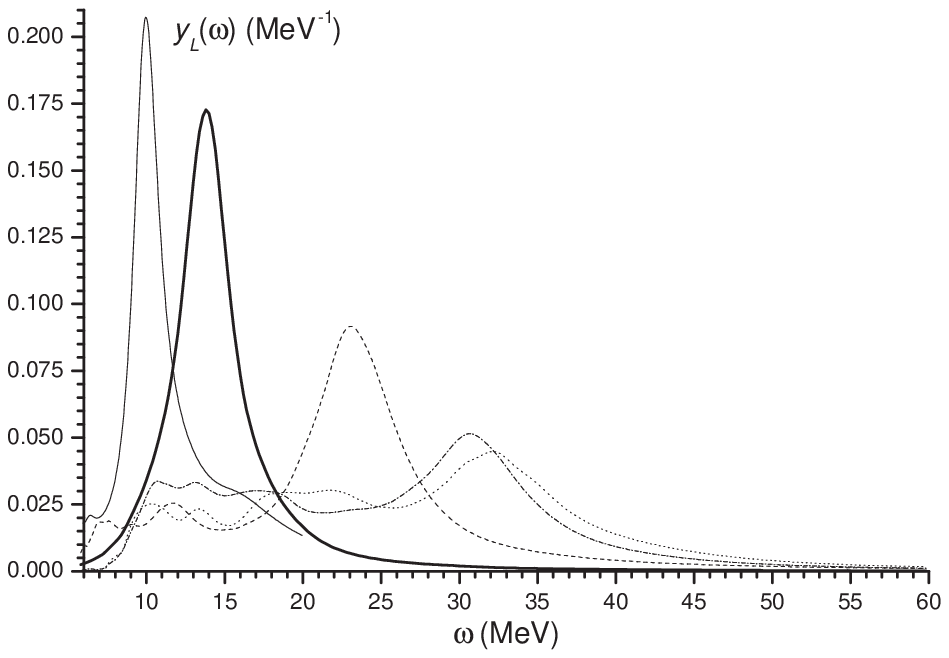}
\caption{The calculated relative strength function for isoscalar GRs
in $^{208}$Pb. The thick, thin, dashed, dotted, and dash-dotted lines are
for the ISGMR, ISGQR, ISGDR, ISGMR2, and ISGQR2, respectively.}
\end{figure}

\section{\label{sec:summary}DISCUSSION OF RESULTS AND SUMMARY}
Within the CRPA-based semi-microscopic approach, all the main properties
of a given GR can be described in a transparent and rather simple way
using universal parameters for medium-heavy and heavy (spherical) nuclei.
The symmetries of the model Hamiltonian are restored via the respective
selfconsistency conditions. In the present version of the approach
the spin-orbit part of the nuclear mean field is mainly responsible for
incomplete restoration of translation invariance of the model. If this part
was taken equal to zero, the SS would exhaust more than 99.5\%
of the respective sum rule, while the gross properties of isoscalar GRs are
practically not changed. The selfconsistency of the present version of
the approach can apparently be improved, provided that the isoscalar
spin-dependent part of the Landau-Migdal particle-hole interaction is taken
into account. The respective study is outside the scope of the present work
and will be addressed in a future publication.

Discussing the results, we start from the gross properties of the isoscalar
GRs. Taking $^{208}$Pb as an example, one can see from the results of the
semi-microscopic calculations of the relative energy-weighted strength
function $y_L(\omega)$ (Fig.~7) the general tendency for changing isoscalar
strength distribution with increasing excitation energy. The main
components of the ISGMR2 and ISGQR2 are not well-collectivized exhausting
a not-too-large part of the respective total strength (Table~\ref{tab:XII}).
As a result, the A-dependence of $\omega^{HE}_{peak}$ for these overtones
is not so regular, as it takes place for more collective GRs
(Table~\ref{tab:I}). According to the data shown in
Tables~\ref{tab:I},~\ref{tab:IV},~\ref{tab:V} the experimental energies of
the ISGMR, ISGQR, and HE-ISGDR in nuclei from a wide mass interval are
reasonably described within the approach. The energy of the LE-ISGDR in
the same nuclei is described satisfactorily (Table~\ref{tab:I},~\ref{tab:IV}).
The energy of the recently found ISGQR2 in $^{208}$Pb $26.9 \pm 0.7$ MeV
\cite{Hun1,Hun2} is also satisfactorily reproduced in the calculations
(Table~\ref{tab:I}). Thus, the calculated energies of the second isoscalar
giant quadrupole and monopole resonances in nuclei under consideration can be
used as a guide in search of these GRs experimentally.

\begin{table}
\caption{\label{tab:VII}Calculated partial branching ratios for direct nucleon
decay of the HE-ISGRs in $^{90}$Zr $(S_\mu = 1)$. Branching ratios are given in \%.} 
\begin{ruledtabular}
\begin{tabular}{rx{2.1}x{2.1}x{2.1}}
$\mu^{-1}$ & \mc{$\bar b^{L=1}_\mu$} & \mc{$\bar b^{L=0}_\mu$} & \mc{$\bar b^{L=2}_\mu$} \\
           & \mc{(18-32 MeV)}        & \mc{(23-39 MeV)}        & \mc{(25-40 MeV)}   \\
\hline
neutron&&&\\
(9/2)$^+$   & 10.2 & 7.6 & 14.1 \\
(1/2)$^-$   & 2.8  & 3.6 & 2.1  \\
(5/2)$^-$   & 4.6  & 5.8 & 5.0  \\
(3/2)$^-$   & 5.7  & 7.3 & 4.8  \\
(7/2)$^-$   & 4.5  & 7.2 & 7.2  \\ 
(1/2)$^+$   & 0.9  & 2.4 & 1.8  \\
\hline
$\bar b^{tot}_n$ & 29.6 & 40.0 & 39.9 \\
\hline
proton&&&\\
(1/2)$^-$   & 4.0 & 4.0 & 3.0 \\
(3/2)$^-$   & 8.0 & 8.2 & 6.4 \\
(5/2)$^-$   & 5.1 & 7.3 & 6.4 \\
(7/2)$^-$   & 4.0 & 8.6 & 7.7 \\
(1/2)$^+$   & 1.7 & 3.3 & 2.6 \\
(3/2)$^+$   & 0.8 & 3.2 & 1.9 \\
\hline
$\bar b^{tot}_p$ & 24.0 & 39.5 & 31.4 \\
\end{tabular}
\end{ruledtabular}
\end{table}
\begin{table}
\caption{\label{tab:VIII} Calculated partial branching ratios for direct
nucleon decay of the HE-ISGRs in $^{116}$Sn ($S_\mu = 1$ and $S_\mu=v^{2}_\mu$
are used in calculations of proton and neutron branching ratios,
respectively). Branching ratios are given in \%.} 
\begin{ruledtabular}
\begin{tabular}{rx{1.3}x{2.2}x{2.2}x{2.2}}
$\mu^{-1}$ & \mc{$S_\mu$} & \mc{$\bar b^{L=1}_\mu$} &
             \mc{$\bar b^{L=0}_\mu$} & \mc{$\bar b^{L=2}_\mu$} \\
           &              & \mc{(15-35 MeV)}        &
             \mc{(22-38 MeV)}        &  \mc{(25-37 MeV)}  \\
\hline
neutron&&&&\\
(1/2)$^-$  & 0.006 & 0.04 & 0.20 & 0.06 \\
(3/2)$^-$  & 0.007 & 0.07 & 0.27 & 0.07 \\
(7/2)$^-$  & 0.011 & 0.11 & 0.16 & 0.06 \\
(11/2)$^-$ & 0.188 & 1.9  & 0.83 & 2.7  \\
(3/2)$^+$  & 0.195 & 0.84 & 0.76 & 0.45 \\
(1/2)$^+$  & 0.362 & 0.89 & 1.0  & 0.46 \\
(7/2)$^+$  & 0.862 & 4.65 & 3.7  & 4.75 \\
(5/2)$^+$  & 0.896 & 6.7  & 5.9  & 4.4  \\
(9/2)$^+$  & 0.992 & 4.9  & 5.8  & 8.7  \\
(1/2)$^-$  & 0.994 & 1.8  & 2.3  & 1.4  \\
(3/2)$^-$  & 0.996 & 3.45 & 4.8  & 3.2  \\
\hline
\multicolumn{2}{c}{$\bar b^{tot}_n$} & 28.8 & 25.7 & 34.7 \\
\hline
proton&&&&\\
(9/2)$^+$ & & 4.5 & 8.2 & 9.0 \\
(1/2)$^-$ & & 2.2 & 3.6 & 1.9 \\
(3/2)$^-$ & & 3.9 & 7.3 & 4.0 \\
(5/2)$^-$ & & 1.8 & 5.3 & 2.5 \\
(7/2)$^-$ & & 1.1 & 5.6 & 3.1 \\
(1/2)$^+$ & & 0.2 & 1.8 & 1.4 \\
(3/2)$^+$ & & 0.1 & 1.3 & 0.4 \\
(5/2)$^+$ & &     & 1.7 &  0.6 \\
\hline
\multicolumn{2}{c}{$\bar b^{tot}_p$} & 14.0 & 35.0 & 22.9\\
\end{tabular}
\end{ruledtabular}
\end{table}

We try to elucidate the universal phenomenological description of the total
width of an arbitrary GR using an appropriate smearing parameter with the
saturation-like energy dependence of Eq.~(\ref{iomega}). In applying to
isovector GRs, a similar attempt was found to be satisfactory \cite{Rodin}.
According to this description, the total width of low-energy GRs (except for
the isobaric analog resonance) is mainly due to the spreading effect.
Such a case is realized for the main-tone isoscalar monopole and quadrupole
resonances. As follows from CRPA calculations, a significant part of the total
width of high-energy GRs is due to the particle-hole strength distribution and
coupling to the continuum. The rest is due to the spreading effect,
which leads also to averaging the strength distribution over the energy.
Regarding the experimental total widths of the isoscalar GRs in nuclei under
consideration, we note that the widths of the ISGMR and ISGQR, the RMS energy
dispersion for both components of the ISGDR are well described within the
approach (Tables~\ref{tab:I},~\ref{tab:V}). The experimental total width of
the ISGQR2 in $^{208}$Pb $6.0 \pm 1.3$ MeV \cite{Hun1} agrees well with the RMS
energy dispersion, calculated for the main component of this resonance
(Table~\ref{tab:XII}). It is worth to note the noticeable scatter in the
experimental data on the total width of the HE-ISGDR \cite{Itoh,Hun1,Hun2}.
This scatter and also the scatter of experimental relative strengths $x_L$ are
apparently explained by uncertainties in the substraction of the underlying
continuum in the analysis of the $(\alpha, \alpha')$-reaction cross section.
For this reason we do not compare in the present work the calculated $x_L$
values (Tables~\ref{tab:II} and \ref{tab:III}) with the respective experimental
data. Nevertheless, one can compare the calculated (Fig.~3b) and experimental
(Ref.~\cite{Nayak}) strength functions $y_{L=1}^{(2)}(\omega)$ for $^{58}$Ni
and discover satisfactory agreement.

As expected, the radial dependence of the one-node transition density
$R_L(r,\omega_{peak})$ for the ISGMR and HE-ISGDR (Figs.~1b and 3c) is rather
close to that of the respective transition density calculated within the
scaling model \cite{String}, provided the ground-state density of
Eq.~(\ref{grstrho}) is used in the calculations. It can be seen, for instance,
from Fig.~2 of Ref.~\cite{Gor2}. We note also, that the difference of the
transition densities calculated at the peak energy of each ISGDR component is
not so large (Figs.~3b and 3c) to allow for searching for exotic explanations
of LE-ISGDR nature. As it is also expected, the overtone transition density has
one extra node inside the nucleus relative to the main-tone transition density
(Figs.~1b,~2b,~3c,~5b,~6b). Concluding consideration of the gross properties of
the isoscalar GRs, we note that the use of the modified smearing procedure
(as compared with that of Ref.\cite{Gor2}) leads practically to the same
results. It can be seen from comparison of the data from Tables~\ref{tab:II}
and \ref{tab:III} with those from Tables~I and II of Ref.~\cite{Gor2},
respectively.

The use of the modified smearing procedure (Subsect.~\ref{sec:smear})
allows us to describe within the present approach the direct-nucleon-decay
branching ratios for high-energy GRs. The respective calculation results
obtained for the isoscalar overtones in nuclei under consideration, are given
in Tables~\ref{tab:VI}-\ref{tab:X}. For the ISGDR, the total
direct-nucleon-decay branching ratio decreases with decrease of the peak
energy (and with increase of the mass number). The relative change is larger for
the total proton branching ratios due to the difference in the penetrability
factors. We note, that the results of Ref.~\cite{Gor3}, where the main
shortcoming in evaluation of the direct-nucleon-decay branching ratios for
high-energy GRs was eliminated (Subsect.~\ref{sec:smear}), are close to those
of Tables~\ref{tab:VI}-\ref{tab:XI}.

\begin{table}
\caption{\label{tab:IX} Calculated partial branching ratios for direct
neutron decay of the HE-ISGRs in $^{144}$Sm $(S_\mu=1)$. Branching ratios
are given in \%.} 
\begin{ruledtabular}
\begin{tabular}{rx{1.1}x{1.1}x{1.1}}
$\mu^{-1}$ & \mc{$\bar b^{L=1}_\mu$} & \mc{$\bar b^{L=0}_\mu$} & \mc{$\bar b^{L=2}_\mu$} \\
           & \mc{(15-35 MeV)}        & \mc{(21-38 MeV)}        & \mc{(24-36 MeV)}   \\
\hline
(1/2)$^+$   &  1.9 & 2.0 & 1.0 \\
(3/2)$^+$   &  3.4 & 3.3 & 2.2 \\
(11/2)$^-$  &  5.2 & 4.9 & 9.9 \\
(5/2)$^+$   &  5.1 & 5.7 & 4.4 \\
(7/2)$^+$   &  2.6 & 4.1 & 3.7 \\ 
(9/2)$^+$   &  1.8 & 4.3 & 5.2 \\
\hline
$\bar b^{tot}_n$ & 23.1 & 35.1 & 35.3 \\
\end{tabular}
\end{ruledtabular}
\end{table}
\begin{table}
\caption{\label{tab:X} Calculated partial branching ratios for direct
nucleon decay of the HE-ISGRs in $^{208}$Pb $(S_\mu=1)$. Branching ratios
are given in \%.} 
\begin{ruledtabular}
\begin{tabular}{rx{1.1}x{1.1}x{1.1}}
$\mu^{-1}$ & \mc{$\bar b^{L=1}_\mu$} & \mc{$\bar b^{L=0}_\mu$} & \mc{$\bar b^{L=2}_\mu$} \\
           & \mc{(15-35 MeV)}        & \mc{(25-35 MeV)}        & \mc{(25-35 MeV)}   \\
\hline
neutron&&&\\
(1/2)$^-$   & 0.7 & 0.2 & 0.2 \\
(5/2)$^-$   & 2.5 & 0.7 & 1.3 \\
(3/2)$^-$   & 1.8 & 0.6 & 0.6 \\
(13/2)$^+$  & 3.8 & 1.3 & 7.5 \\
(7/2)$^-$   & 4.3 & 1.9 & 2.7 \\ 
(9/2)$^-$   & 2.1 & 1.1 & 3.0 \\
\hline
$\bar b^{tot}_n$ & 22.4 & 24.7 & 32.5\\
\hline
proton&&&\\
(1/2)$^+$   & 1.2  & 2.6 & 1.2 \\
(3/2)$^+$   & 1.4  & 4.4 & 2.2 \\
(11/2)$^-$  & 0.5  & 6.2 & 2.7 \\
(5/2)$^+$   & 1.4  & 7.2 & 3.5 \\
(7/2)$^+$   & 0.28 & 3.1 & 0.8 \\
(9/2)$^+$   & 0.07 & 2.8 & 0.7 \\
\hline
$\bar b^{tot}_p$ & 4.9 & 31.5 & 13.0 \\
\end{tabular}
\end{ruledtabular}
\end{table}
\begin{table}
\caption{\label{tab:XI} Partial branching ratios for direct proton decay of
the HE-ISGRs in $^{208}$Pb into some one-hole states of $^{207}$Tl.
Experimental spectroscopic factors $S_\mu$ taken from Ref.~\cite{Bobel}
are used in calculations. Excitation-energy intervals are taken the same
as in Table~\ref{tab:X}. Calculation results for decays of the HE-ISGDR
are compared with the experimental data of Refs.~\cite{Hun2,Garg}.
Branching ratios are given in \%.}
\begin{ruledtabular}
\begin{tabular}{rx{1.2}x{1.2}cccx{1.2}x{1.2}}
\multicolumn{1}{r}{\begin{tabular}{r}$\mu^{-1}$\end{tabular}} &
   \mc{$S_\mu$} & \mc{$\bar b^{L=1}_\mu$} & & \cite{Hun2} &
   \cite{Garg} \footnote{Preliminary results} &
   \mc{$\bar b^{L=0}_\mu$} & \mc{$\bar b^{L=2}_\mu$} \\
\hline
\multicolumn{1}{r}{\begin{tabular}{r} (1/2)$^+$ \\ (3/2)$^+$ \end{tabular}} &
\mc{\begin{tabular}{x{1.2}} 0.55 \\ 0.57 \end{tabular}} &
\mc{\begin{tabular}{x{1.2}} 0.65 \\ 0.80 \end{tabular}} &
\Bigg\} & $2.3\pm1.1$ &
\mc{\begin{tabular}{c} $0.34\pm0.06$ \\ $0.61\pm0.10$ \end{tabular}} &
\mc{\begin{tabular}{x{1.2}} 1.43 \\ 2.51 \end{tabular}} &
\mc{\begin{tabular}{x{1.2}} 0.66 \\ 1.25 \end{tabular}} \\
\multicolumn{1}{r}{\begin{tabular}{r} (11/2)$^-$ \\ (5/2)$^+$ \end{tabular}} &
\mc{\begin{tabular}{x{1.2}} 0.58 \\ 0.54 \end{tabular}} &
\mc{\begin{tabular}{x{1.2}} 0.29 \\ 0.75 \end{tabular}} &
\Bigg\} & $1.2\pm0.7$ &
\mc{\begin{tabular}{c} $0.31\pm0.05$ \\ $1.07\pm0.17$ \end{tabular}} &
\mc{\begin{tabular}{x{1.2}} 3.60 \\ 3.89 \end{tabular}} &
\mc{\begin{tabular}{x{1.2}} 1.57 \\ 1.89 \end{tabular}} \\
\multicolumn{1}{r}{\begin{tabular}{r}(7/2)$^+$\end{tabular}} &
0.26 & 0.02 & & & & 0.81 & 0.21 \\
\hline
\multicolumn{2}{c}{$\sum\bar b_\mu^L$} &
2.51 & & & & 12.24 & 5.58 \\
\end{tabular}
\end{ruledtabular}
\end{table}

The recent experimental data of Refs.~\cite{Hun2,Garg} on partial
direct-proton-decay branching ratios for the HE-ISGDR in $^{208}$Pb are
satisfactorily described within the present approach, provided that experimental
spectroscopic factors for the final single-hole states of $^{207}$Tl
(Table~\ref{tab:XI}) are taken into account. In Ref.~\cite{Hun2} direct neutron
decay of the same resonance into the final states of $^{207}$Pb from
excitation-energy interval 0--6 MeV has been also observed. The deduced
branching ratio $23 \pm 5$\%
reasonably agrees with the value 15.2\%
obtained with the use of unit spectroscopic factor for the respective
one-hole states (Table~\ref{tab:X}). Reasonable description of the above data
allows us to infer that the calculation results shown in
Tables~\ref{tab:VI}-\ref{tab:IX} will be useful for the analysis of forthcoming
experimental data on direct nucleon decays of the HE-ISGDR in several
medium-heavy and heavy nuclei \cite{Fuji}. Some evidence for direct proton
decay of the ISGQR2 in $^{208}$Pb in the coincidence $(\alpha, \alpha' p)$
experiments have been recently found \cite{Hun1,Hun2}. It allows us to hope
that the calculated parameters of the ISGQR2 and ISGMR2 (including the energy,
total width, transition density, and branching ratios for direct-nucleon-decay)
would be also useful for experimental search of these resonances.

In conclusion, we extend a CRPA-based partially selfconsistent
semi-microscopic approach to describe direct nucleon decay of high-energy
giant resonances. The main properties of the isoscalar overtones
(ISGDR, ISGMR2, ISGQR2) in a few singly- and doubly-closed-shell nuclei are
described within the approach and found to be in reasonable agreement with
available experimental data including the latest ones. Abilities of the
approach for description of the main-tone resonances are successfully checked.
Predictions concerning forthcoming experimental data on the isoscalar
overtones are also presented.

\begin{table}
\caption{\label{tab:XII} Paramaters of the ISGMR2 and ISGQR2 calculated for
a certain excitation-energy interval. All the parameters are given in MeV
except for $x$, which is given in \%.} 
\begin{ruledtabular}
\begin{tabular}{rcx{2.1}x{1.1}ccx{2.1}x{1.1}c}
& \multicolumn{4}{c}{ISGMR2} & \multicolumn{4}{c}{ISGQR2} \\
nucl. & $\omega_1-\omega_2$ & \mc{$\bar\omega$} & \mc{$\Delta$} & $x$
      & $\omega_1-\omega_2$ & \mc{$\bar\omega$} & \mc{$\Delta$} & $x$ \\
\hline
$^{58}$Ni  & 23-40 & 31.1 & 4.7 & 52 & 25-40 & 32.5 & 4.1 & 45 \\
$^{90}$Zr  & 23-39 & 30.6 & 4.4 & 51 & 25-40 & 32.2 & 4.1 & 48 \\
$^{116}$Sn & 22-38 & 29.6 & 4.6 & 50 & 25-37 & 31.3 & 3.3 & 40 \\
$^{144}$Sm & 21-38 & 29.2 & 4.7 & 56 & 24-36 & 30.4 & 3.3 & 43 \\
$^{208}$Pb & 25-35 & 30.2 & 2.8 & 34 & 25-35 & 30.0 & 2.6 & 38 \\
\end{tabular}
\end{ruledtabular}
\end{table}

\begin{acknowledgments}
The authors are indebted to M.~Fujiwara, U.~Garg, and M.N.~Harakeh for
providing the latest experimental data. The authors are thankful to
M.N.~Harakeh for many interesting discussions and valuable remarks concerning
the manuscript. Two authors (I.V.S. and M.H.U.) are grateful to U.~Garg for
hospitality during their stay at University of Notre Dame and acknowledge
support from the National Science Foundation under grant No.~PHY-0140324. 
M.H.U. is grateful to M.N.~Harakeh for hospitality during the long-term visit
at KVI and acknowledges support from the ``Nederlandse organisatie voor
wetenschappelijk onderzoek'' (NWO).
\end{acknowledgments}

\end{document}